\documentstyle[12pt]{article}
\begin{document}
\large
\rightline{UT-687,1994}
\vfil
\vfil
\Large
\begin{center}
 A Vector-like Extension of the Standard Model
\\
\vfil
   Kazuo Fujikawa
\\

Department of Physics, University of Tokyo\\
         Bunkyo-ku, Tokyo  113, Japan
\vfil
\vfil
Abstract

\end{center}
\large
A vector-like extension of the standard model for heavier
quarks and leptons with $SU(2)\times U(1)$ gauge symmetry and only one
Higgs doublet is examined. This scheme incorporates infinitely many
fermions and  avoids the appearance of a strongly
interacting sector.
The model is perturbatively well controllable, and
constraints such as $b\rightarrow s\gamma$ and GIM suppression in
general are naturally satisfied if the on-set of
heavier fermion mass scale is chosen
at a few TeV. In this scheme, the physical Higgs particle
generally mediates leptonic as well as quark flavor-changing processes
at a rate below the present experimental limit.
Vector-like models, if suitably
defined, provide a viable and interesting scheme for physics beyond 1 TeV.
\\
\\
(to be published in Prog. of Theor. Phys.)
\newpage
\section{Introduction}

\par
     The success of the standard model[1] of quarks and leptons is
impressive. Perhaps the top quark,whose mass was constrained to be
in a narrow range by experiments at LEP, has been discovered at
Fermilab[2]. The weak coupling of
the top quark is predicted
to be almost purely left-handed on the basis of the $\rho$-parameter
analysis[3] and also by an analysis of the $b\rightarrow s\gamma$ decay[4].
At the same time, an abrupt end of  proliferation of quarks
and leptons(in particular, light neutrinos) at the 3rd generation
is rather mysterious.
     One of the reasons why the generations with heavier  masses
are prohibited may be the dynamical stability of  Weinberg-Salam
theory with its chiral (left-handed) weak couplings. Some time ago,
we performed
an analysis of this problem of heavy fermions[5],although
at that time the top quark was widely believed to be not so heavy
 as the present day experiments indicate.
We here resume this analysis .

In the standard model, all the masses of
gauge bosons and fermions are generated by the Higgs
mechanism.
For this reason, the successful physics of 3 generations of
light fermions below 1 TeV  may be called as the " Higgs world".
In the following,
we  discard the dynamics of Higgs mechanism for a moment,
and we  regard the known leptons and quarks
together with W and Z bosons as all massless.
 By this extreme simplification of the picture,
we can think about the global aspects of fermion spectrum better.
We then examine whether
heavier quarks and leptons with masses of  TeV region, for example,
can be accommodated in the standard model.

     We first recapitulate the essence of the arguments in Ref.[5].
The first observation is a consequence of the coupling of  W boson
to the fermion doublet $ \psi_{k},k=1,2$ generically defined by
\begin{equation}
{\cal
L}=(1/2)g\bar{\psi}_{k}(T^{a})_{kl}\gamma^{\mu}(a+b\gamma_{5})\psi_{l}W_{\mu}^{a}
\end{equation}
The longitudinal coupling of $W_{\mu}^{a}$, which is related to Higgs
mechanism , is studied by replacing $W_{\mu}^{a}\rightarrow (2/gv)\partial
_{\mu}S^{a}(x)$ in (1) with $ S^{a}(x)$ the unphysical Higgs scalar.
We then obtain
\begin{equation}
{\cal L}=a(\frac{m_{k}-m_{l}}v)\overline{\psi}_{k}(T^{a})_{kl}\psi_{l}S^{a}

+b(\frac{m_{k}+m_{l}}v)\overline{\psi}_{k}(T^{a})_{kl}\gamma_{5}\psi_{l}S^{a}
\end{equation}
by using the equations of motion.

The typical mass scale of the Higgs world is
\begin{equation}
v=247GeV
\end{equation}
and thus
\begin{equation}
   b(m_{k}+m_{l})\gg v  ,\ or ,\   a|m_{k}-m_{l}|\gg v \\
\end{equation}
induces a strongly interactiong sector into the standard model.
We here regard
the situation in (4) as unnatural.In other words, the standard model
with chiral couplings does not accommodate  fermions with masses much
larger than the Higgs scale (3).

Another ingredient of the arguments in Ref.[5] is provided by
the one-loop effective potential[6]

$V(\varphi)=-(\frac{1}2)M^{2}\varphi^{2} + f\varphi^{4}$
\begin{equation}
          +(64\pi^{2})^{-1}Tr\{3\mu_{\varphi}^{4}\ln\mu_{\varphi}^{2}
+ M_{\varphi}^{4}\ln M_{\varphi}^{2} - 4m_{\varphi}^{4}\ln m_{\varphi}^{2}\}
\end{equation}
where $\mu_{\varphi}$,$M_{\varphi}$ and $m_{\varphi}$ are ,respectively,
the zeroth-order vector ,scalar and spinor mass matrices for a scalar
field vacuum expectation value $\varphi$. When one has fermions with
mass much larger than (3), the stability of the potential (5) suggests that
\begin{equation}
m_{H} \geq \sqrt{2}m_{F} \gg M_{W}
\end{equation}
Namely, the Higgs mass is forced to become heavy by the appearance of
heavy fermions, and the heavy (physical) Higgs in turn induces a
strongly interacting sector into  Weinberg-Salam theory[7]. We again
regard this situation unnatural. It is possible to make the analysis
in (6) more precise, but the semi-quantitative analysis (6) is
sufficient for our present purpose.

The heavier fermions in the standard scheme, if they should exist without
spoiling dynamical stability in a perturbative sense, should thus
have almost pure vector-like couplings (i.e., $b \simeq 0 $ in (4))
and that the mass of the fermion doublet should be almost degenerate
(i.e., $a|m_{k}-m_{l}| \leq v$ in (4)).This latter constraint is also
desirable to ensure that the $\rho$-parameter does not receive
unacceptably large radiative corrections[2].
If  heavier quarks and leptons satisfy the above conditions, they
have no sizable couplings to the Higgs sector. In other words, their
masses primarily come from dynamics
which is different from  the Higgs mechanism in the standard model, and
consequently,
the stability argument on the basis of the
effective potential (5) becomes irrelevant for these heavier fermions.
Also, the breaking mechanism of SU(2) (both of local as well as custodial)
in the standard model
is concluded to be a typical phenomenon in the energy scale of $v$.

\section{A Vector-like Extention of the Standard Model}
\par
The present note is mainly motivated by an analysis in Ref.[4] and the
appearance of a stimulating
scheme which
satisfies  main features of the  constraints discussed above. This scheme
was introduced in the name of generalized
Pauli-Villars regularization[8]. We regard the fermion sector
of the generalized Pauli-Villars regularization as a model of
realistic
fermion spectrum, and thus  unphysical bosonic Dirac fields appearing
in the regularization are
of course excluded. A crucial property of the Pauli-Villars
regularization
is that  heavier fermions, when their masses become large, decouple
from the world of light chiral fermions.
In the considerations so far\footnote{
A somewhat related scheme was
also considered by K. Inoue from  different considerations[9].},
the masses of fermions other than the conventional leptons
and quarks
are taken to be of the order of the grand unification mass scale
or the Planck mass. In this respect, our physical motivation
is completely different.

To be specific , we consider an $SU(2){ \times} U(1)$ gauge theory
written in an abbreviated notation
\begin{equation}
{\cal L}_{L}=\overline{\psi}i\gamma^{\mu}D_{\mu}\psi
            - \overline{\psi}_{R}M\psi_{L}
            - \overline{\psi}_{L}M^{\dagger}\psi_{R}
\end{equation}
with
\begin{equation}
\not{\!\! D}=\gamma^{\mu}(\partial_{\mu} - igT^{a}W_{\mu}^{a}
            - i(1/2)g^{\prime}Y_{L}B_{\mu})
\end{equation}
and  $Y_{L}=1/3$ for quarks and $Y_{L}=-1$ for leptons. The field
$\psi$ in (7) is a column vector consisting of an infinite number of
$SU(2)$ doublets, and
the infinite dimensional $nonhermitian$ mass matrix $M$ satisfies the
index condition
\begin{equation}
\dim\ker(M^{\dagger}M) = 3,\ \dim\ker(M M^{\dagger})=0
\end{equation}
In the  explicit "diagonalized" expression of $M$
\begin{eqnarray}
M&=&\left(\begin{array}{ccccccc}
          0&0&0&m_{1}&0    &0    &..\\
          0&0&0&0    &m_{2}&0    &..\\
          0&0&0&0    &0    &m_{3}&..\\
          .&.&.&.    &.    &.    &..
          \end{array}\right)\nonumber\\
M^{\dagger}M&=&\left(\begin{array}{cccccc}
          0&&&&&                 \\
           &0&&&0&               \\
           &&0&&&                \\
           &&&m_{1}^{2}&&        \\
           &0&&&m_{2}^{2}&       \\
           &&&&&..
          \end{array}\right)\nonumber\\
M M^{\dagger}&=&\left(\begin{array}{cccccc}
           m_{1}^{2}&&&&&            \\
                    &m_{2}^{2}&&0&&   \\
                    &&m_{3}^{2}&&&   \\
                    &0&&..&&         \\
                    &&&&..&
          \end{array}\right)
\end{eqnarray}
the fermion $\psi$ is written as
\begin{equation}
 \psi_{L}=(1-\gamma_{5})/2\left(
 \begin{array}{c}
  \psi_{1}\\ \psi_{2}\\ \psi_{3}\\ \psi_{4}\\.
 \end{array}
 \right), \ \
 \psi_{R}=(1+\gamma_{5})/2\left(
 \begin{array}{c}
  \psi_{4}\\ \psi_{5}\\ \psi_{6}\\.\\.
 \end{array}
 \right)
\end{equation}
We thus have 3 massless left-handed $SU(2)$ doublets $\psi_{1},\psi_{2},
\psi_{3}$, and an
infinite series of vector-like massive $SU(2)$ doublets $\psi_{4},
\psi_{5},...$ with
masses $m_{1},m_{2},..$ as is seen in\footnote{
One may introduce \underline{constant} complete orthonormal sets
$\{ u_{n} \}$ and $\{ v_{n}\}$ defined by \\
$M^{\dagger}Mu_{n} = 0$ for $n=-2, -1, 0$,\\
$M^{\dagger}Mu_{n} = m_{n}^{2}, M M^{\dagger}v_{n} = m_{n}^{2}v_{n}$
for $n = 1, 2, ...$\\
by assuming the index condition (9). One then has $Mu_{n} = m_{n}v_{n}$
for $m_{n}{\neq} 0$  by choosing the phase of $v_{n}$
and $Mu_{n} = 0$ for $m_{n}=0$ . When one expands\\
$\psi_{L}= \sum_{n=-2}^{\infty} \psi_{n+3}^{L}u_{n},
 \psi_{R}= \sum_{n=1}^{\infty} \psi_{n+3}^{R}v_{n}$\\
one recovers the mass matrix (10) and the relation (12).}

\begin{eqnarray}
{\cal L}_{L}&=&\bar{\psi}_{1}i\not{\!\! D}(\frac{1-\gamma_{5}}{2})\psi_{1}
               +\bar{\psi}_{2}i\not{\!\! D}(\frac{1-\gamma_{5}}{2})\psi_{2}
                \nonumber\\
            & &+\bar{\psi}_{3}i\not{\!\! D}(\frac{1-\gamma_{5}}{2})\psi_{3}
                \nonumber\\
            & &+\bar{\psi}_{4}(i\not{\!\! D} -m_{1})\psi_{4}
               +\bar{\psi}_{5}(i\not{\!\! D} -m_{2})\psi_{5} + ...
\end {eqnarray}

An infinite number of right-handed fermions in a doublet notation are also
introduced by( again in an abbreviated notation)
\begin{equation}
{\cal L}_{R}=\overline{\phi}i\gamma^{\mu}(\partial_{\mu}-i(1/2)g^{\prime}
Y_{R}B_{\mu})\phi - \overline{\phi}_{L}M^{\prime}\phi_{R}
-\overline{\phi}_{R}(M^{\prime})^{\dagger}\phi_{L}
\end{equation}
where
\begin{equation}
Y_{R}=\left(\begin{array}{cc}
            4/3&0\\
            0&-2/3
            \end{array}\right)
\end{equation}
for quarks and
\begin{equation}
Y_{R}=\left(\begin{array}{cc}
            0&0\\
            0&-2
            \end{array}\right)
\end{equation}
for leptons, and the mass matrix $M^{\prime}$   satisfies the index
condition
(9) but in general it may have different mass eigenvalues from
those in(10). After the diagonalization of $M^{\prime}$, $\phi$ is
written as
\begin{equation}
 \phi_{L}=(1-\gamma_{5})/2\left(
 \begin{array}{c}
  \phi_{4}\\ \phi_{5}\\ \phi_{6}\\ .\\ .
 \end{array}
 \right), \ \
 \phi_{R}=(1+\gamma_{5})/2\left(
 \begin{array}{c}
  \phi_{1}\\ \phi_{2}\\ \phi_{3}\\ \phi_{4}\\ .
 \end{array}
 \right)
\end{equation}
Here, $\phi_{1}, \phi_{2}$,and  $ \phi_{3}$ are right-handed and massless,
and $\phi_{4}, \phi_{5},....$ have masses $m_{1}^{\prime}, m_{2}^{\prime}$,..
\begin{eqnarray}
{\cal L}_{R}&=&\bar{\phi}_{1}i\not{\!\! D}(\frac{1+\gamma_{5}}{2})\phi_{1}
               +\bar{\phi}_{2}i\not{\!\! D}(\frac{1+\gamma_{5}}{2})\phi_{2}
                \nonumber\\
            & &+\bar{\phi}_{3}i\not{\!\! D}(\frac{1+\gamma_{5}}{2})\phi_{3}
                \nonumber\\
            & &+\bar{\phi}_{4}(i\not{\!\! D} -m_{1}^{\prime})\phi_{4}
               +\bar{\phi}_{5}(i\not{\!\! D} -m_{2}^{\prime})\phi_{5} + ...
\end {eqnarray}
with
\begin{equation}
   \not{\!\! D}= \gamma^{\mu}(\partial_{\mu}-i(1/2)g^{\prime}
                 Y_{R}B_{\mu})
\end{equation}

The present model is vector-like and manifestly anomaly-free
before the  breakdown  of parity (9);after the breakdown of
parity,the model still stays anomaly-free provided that both of $M$ and
$M^{\prime}$ satisfy the index condition (9). In this scheme, the anomaly
 is
caused by the left-right asymmetry, in particular, in the sector of
(infinitely) heavy fermions; in this sense, the parity breaking (9)
may be termed "hard breaking". Unlike  conventional vector-like
models with a finite number of components[10], the present scheme avoids
the
appearance of a strongly interacting right-handed sector despite of the
presence of heavy fermions.
A truncation of the present scheme to a finite number of heavy
fermions (for example, to only one heavy doublet in $\psi$) is still
consistent, although it is no more called vector-like.

The massless fermion sector in the above scheme  reproduces the same
 set of fermions
as in the standard model. However, heavier fermions have distinct
features. For example, the heavier fermion doublets with the smallest
masses are
described by
\begin{eqnarray}
{\cal L}&=&\overline{\psi}_{4}i\gamma^{\mu}(\partial_{\mu}-igW_{\mu}^{a}
           -i(1/2)g^{\prime}Y_{L}B_{\mu})\psi_{4}-m_{1}\overline{\psi}_{4}
           \psi_{4}\nonumber\\
        & &+\overline{\phi}_{4}i\gamma_{\mu}(\partial_{\mu}
           -i(1/2)g^{\prime}Y_{R}B_{\mu})\phi_{4}
           -m_{1}^{\prime}\overline{\phi}_{4}\phi_{4}
\end{eqnarray}
The spectrum of fermions is thus $doubled$ to be vector-like in the
sector  of heavy fermions and ,at the same time, the masses of $\psi$ and
$\phi$ become non-degenerate, i.e., $m_{1}{\neq}m_{1}^{\prime}$.
As a result, the fermion number anomaly[11]
is generated only
by the first 3 generations of light fermions;the violation
of baryon number is not enhanced by the presence of heavier fermions.
The masses  of  heavy doublet components in $\psi$  are degenerate
 in the present zeroth order approximation. The masses of heavy doublets in
$\phi$ have nothing to do with custodial SU(2) in the zeroth order
approximation, but they are taken to be degenerate for simplicity.

In the present scheme we distinguish two classes of chiral symmetry
breaking;one which is
related to the breaking of gauge symmetry (Higgs mechanism), and the
other which is related to the mass of heavier fermions but not related to
the breaking of gauge symmetry. The transition from one class of
chiral symmetry breaking to the other, which is also accompanied by the
transition from chiral to vector-like gauge couplings, is  assumed
to take place at the mass scale of the order of $v$ in (3). In any case
if the $SU(2){\times}U(1)$ gauge symmetry should be universally valid
regardless of the magnitude of the mass of fermions, just like
electromagnetism and gravity, the coupling of heavier fermions is
required to become vector-like:\ Heavy gauge bosons can naturally
couple to light fermions, but the other way around imposes a
stringent constraint on the chirality of fermions.
We are  here interested
in the possible on-set of heavier fermions at the order of a few TeV,
although these vector-like components are often assumed to acquire
masses of the order of grand unification scale(Georgi's survival
hypothesis[12]).

\section{Light Fermion Masses and Higgs Mechanism}

\par
As for the mass generation of the first 3 generations of quarks and
leptons and also the custodial $SU(2)$ breaking of heavier fermions,
one  may introduce a Yukawa interaction for quarks, for example, in
an abbreviated notation
\begin{eqnarray}
{\cal L}_{Y}&=&\bar{\psi}_{L}G_{u}\varphi\phi_{R}^{(u)}
         + \bar{\psi}_{L}G_{d}\varphi^{c}\phi_{R}^{(d)}\nonumber\\
            &+&\bar{\psi}_{R}G_{u}^{\prime}\varphi\phi_{L}^{(u)}
         + \bar{\psi}_{R}G_{d}^{\prime}\varphi^{c}\phi_{L}^{(d)} + h.c.
\end{eqnarray}
where  $\varphi(x)$ is the conventional Higgs doublet ( and
$\varphi(x)^{c}$ is its conjugate) , and
$G_{u}, G_{d}, G_{u}^{\prime}$ and $G_{d}^{\prime}$ are
infinite dimensional
coupling matrices acting on $\psi$ and $\phi$. Corresponding to the
presence of only one $W$-boson, we here assume the existence of only
one Higgs doublet.
The fields $\bar{\psi}_{L}$ or $\bar{\psi}_{R}$ in (20) stands for the
doublets in (7), and $\phi_{R}^{(u)}$ (or $\phi_{L}^{(u)}$) and
$\phi_{R}^{(d)}$ (or $\phi_{L}^{(d)}$), respectively, stand for the
upper and lower components of the doublets $\phi$ in (13). If one
retains only the first two terms and their conjugates in (20) and if
only the massless components of $\psi_{L}$ and $\phi_{R}$ in (11)
and (16) are considered, (20) reduces to the Higgs coupling of the
standard model.

We  postulate that the coupling matrices G  ( which generically include
$G^{\prime}$ hereafter)
are such that
the interaction (20) is perturbatively well controllable,namely, the
typical element of coupling matrices $G$ is bounded by the gauge
coupling $g$,
\begin{equation}
    |G|{\leq}g
\end{equation}
By this
way the masses of the first 3 generations of light fermions are generated
from (20) below the mass scale in (3).
For the heavier fermions,
the interaction (20) introduces the breaking of custodial $SU(2)$ and
also  fermion mixing.After the conventional $SU(2)$ breaking,
\begin{equation}
{\langle}\varphi{\rangle}=v/\sqrt{2}
\end{equation}
one may diagonalize the mass matrix
in (20) together with the mass matrices in (7) and (13).This introduces
a generalization of the ordinary fermion mixing matrix [1].
If one assumes a generic situation,
\begin{eqnarray}
   & & m_{i}, m_{j}^{\prime} {\gg} gv, \ \ \ |m_{i}-m_{j}^{\prime}| {\gg}
        gv{\nonumber} \\
   & & |m_{i}-m_{j}| {\gg} gv, \ \ \ |m_{i}^{\prime}-m_{j}^{\prime}| {\gg}
        gv
\end{eqnarray}
for any combination of (renormalized) heavy fermion masses $m_{i}$ and
$m_{j}^{\prime}$,
the masses of heavier fermions are little modified by the Higgs
coupling.
The state mixing between light fermions
and heavier fermions (and also the mixing among heavier fermions)
introduced by the  interaction  (20) is characterized by a dimensionless
parameter
\begin{equation}
  \varepsilon = (1/2)|G|v/m_{i}{\leq}(1/2)gv/m_{i}
   =M_{W}/m_{i}
\end{equation}
As a fiducial value of the on-set of heavy fermion mass, we here choose
\begin{equation}
m_{i} {\sim} a\ few\ TeV
\end{equation}
and thus the dimensionless parameter $\varepsilon$ (24) at
\begin{equation}
\varepsilon {\leq} m_{W}/m_{i} {\simeq} 1/50
\end{equation}

In practical calculations, it is convenient to diagonalize the light
fermion masses in addition to (10) and its analogue of $M^{\prime}$
but leave the mixing of light and
heavy fermions
non-diagonalized, instead of diagonalizing all the masses. In this case
the effects of  heavy fermions on the processes of light fermions are
estimated in the power expansion of $G$. The mass term after diagonalizing
the light quarks in the up-quark sector, for example, is given by
\begin{eqnarray}
(\bar{\psi}_{L},\bar{\Psi}_{L},\bar{\Phi}_{L}){\cal M}\left(
\begin{array}{c}
\phi_{R}\\ \Psi_{R}\\ \Phi_{R}
\end{array}
\right)
+ h.c.
\end{eqnarray}
where the mass matrix ${\cal M}$ is defined by
\begin{eqnarray}
{\cal M}&=& \left(\begin{array}{ccccccccc}
m_{u}&0     &0    &     & & &                   &&\\
0    &m_{c}&0     &     &\Large{0}& &&\tilde{G}v/\sqrt{2}&\\
0    &0     &m_{t}&     & & &                   &&\\
     &     &               &m_{1}&0    &0&                   &&\\
     &\tilde{G}v/\sqrt{2}& &0    &m_{2}&0             &    &Gv/\sqrt{2}&\\
     &                  &  &0    &0    &..            &    &            &\\
     &                  & & &          &  &m_{1}^{\prime}&0            &0\\
     &\Large{0}         & & &Gv/\sqrt{2}& &0             &m_{2}^{\prime}&0\\
     &                  & & &           & &0             &0          &..
\end{array}\right)\nonumber
\end{eqnarray}
To avoid introducing further notational conventions,  we here use the
fields $\psi_{L}$ and $\phi_{R}$ in(27) for the first 3 light
(i.e.,massless in the zeroth order approximation) fermion components of
$\psi$ and $\phi$ in (11) and (16),respectively; $\Psi$ and $\Phi$ stand
for the
remaining heavy quark components of $\psi$ and $\phi$ in (11) and (16).
The coupling matrix $\tilde{G}$ is different from the original $G$ by
the unitary
transformation of light quarks performed in the process of diagonalizing
light quark masses. But the order of magnitude of $\tilde{G}$ is still
the same as that of $G$.

The physical Higgs $H(x)$ coupling in the unitary gauge is given by the
replacement
\begin{equation}
v \rightarrow v + H(x)
\end{equation}
in the above mass matrix(27); this is also true for the light quark masses
$m_{u}, m_{c},$ and $m_{t}$, if one writes ,for example,\\
$m_{u} = (g/2)(m_{u}/m_{W})v $ and $v \rightarrow v + H(x)$.

If one sets $G = 0$ and $\tilde{G} = 0$ in the mass matrix(27), the
light and heavy quark sectors become completely disconnected, not only
in the Higgs coupling but also in the gauge coupling except for the
renormalization effects due to heavy quark loop diagrams. This means
that the \underline{direct} effects of heavy fermions on the processes
involving light quarks and leptons only can be calculated as a power
series in $\tilde{G}$ and $G$, provided that these effects of heavy
fermions are small. We show that these effects are in fact of
controllable magnitude if the condition $|G| \leq g$ in (21) is satisfied
and the mass spectrum of heavy fermions starts at a few TeV. See
eq.(26) for the dimensionless parameter $\varepsilon$.

 We would like to examine some of the
physical implications of the present scheme\footnote{We here assume that
the mass spectrum of heavier fermions is rather sparsely distributed.
We thus estimate the effects of the lightest heavier fermions on
physical processes involving ordinary fermions in the standard model.}.
The influence
of the SU(2) symmetry breaking  which induces the mass splitting of
fermion doublets is small on  heavier
fermions,and  the heavier fermions are relatively stable
against weak decay despite of their large masses. Heavier fermions
are expected to decay mainly into the Higgs particle and light fermions
with natural decay width
\begin{equation}
 \Gamma_{i}{\sim}|G|^{2}m_{i}.
\end{equation}
In this note,we take the Higgs mass at the "natural" value
\begin{equation}
m_{H}\sim v
\end{equation}
with $v$ in (3).

The mass spectrum of light fermions is influenced by heavier fermions
through the mixing in (20). There are basically two kinds of diagrams
shown  in Fig.1 ,which contribute to light quark masses. Fig.1-a gives
rise to a mass correction
\begin{equation}
\sim (|G|v/m_{i})^{2}|G|v \simeq \varepsilon^{2}|G|v
\end{equation}
and Fig.1-b gives
\begin{equation}
\sim (|G|v/m_{i})^{2}m_{l} \simeq  \varepsilon^{2}m_{l}
\end{equation}
with $m_{l}$ the light quark mass. The first contribution (31) is
dominant
for all the light fermions except for the top quark.\\ Numerically,(31)
gives
\begin{equation}
\varepsilon^{2}(|G|v) \leq \varepsilon^{2}m_{W} \leq 40MeV
\end{equation}
if one uses $\varepsilon \leq 1/50$ in (26). The second contribution (32) is
dominant for the
top quark, but numerically
\begin{equation}
\sim \varepsilon^{2}m_{t} \leq 4\times 10^{-4}m_{t}
\end{equation}
and the correction is negligible.

We find it encouraging that the most natural choice $|G| \sim g$ already
gives a sensible result (33),although a certain fine tuning is
required to
account for the actual masses of the electron and up and down quarks.
In this sense, the present model
may be said natural. It may also be interesting to envision a kind of
see-saw picture, namely, the generation of very light fermion
(i.e., electron and up and down quark) masses
primarily from a mixing with heavy fermions.
The typical mass scale of light fermions may then be chosen at around
the center of gravity of light fermion mass spectrum, namely, at $\sim
10 GeV$
 as is indicated in (40) below. If one adopts this picture , the
light fremion masses
appearing in (29) do not represent physical masses.
The physical mass spectrum itself
is, however, an input in the present scheme.

A characteristic feature of the present extension of the standard
model  is that the leptonic as well as quark
flavor is generally violated; this breaking is caused by the mixing of
light and heavy fermions in (20). The diagonalization of mass matrix
does not diagonalize the Higgs coupling in general unlike the standard
model,
and the physical Higgs particle at the tree level
also mediates flavor changing processes although its contribution is not
necessarily a dominant one; in the limit of large heavy fermion masses
$m_{i} \rightarrow \infty$, this flavor changing coupling vanishes.

We first estimate the Higgs and heavy fermion contributions to the
processes which are GIM suppressed in the standard model. The example
we analyze is the $K^{0}-\bar{K^{0}}$ mixing,whose dominant contribution
is given by the Feynman diagram in Fig. 2-a. This diagram gives rise to
an amplitude of the order of\footnote{
The flavor changing processes induced by the state mixing such as
in Fig. 1 without a Higgs particle exchange should be discarded, since
such  processes do not take place if one diagonalizes the entire mass
matrix exactly.}

\begin{eqnarray}
&\sim& (1/4\pi)(|G|^{4}/m_{i}^{2}) \bar{s_{L}}\gamma^{\mu}d_{L}
                                    \bar{s_{L}}\gamma_{\mu}d_{L}
                                                       \nonumber\\
&\simeq& (m_{W}/m_{i})^{2}(|G|/g)^{4}\alpha G_{F}
                                    \bar{s_{L}}\gamma^{\mu}d_{L}
                                    \bar{s_{L}}\gamma_{\mu}d_{L}
\end{eqnarray}
The tree level Higgs exchange in Fig.2-b gives an amplitude of the order
of
\begin{eqnarray}
&\sim& \varepsilon^{4}(|G|/v)^{2} \bar{s_{R}}d_{L}
                                           \bar{s_{R}}d_{L} \nonumber\\
&\simeq& \varepsilon^{4}(|G|/g)^{2}\alpha G_{F}
                                    \bar{s_{R}}d_{L}
                                    \bar{s_{R}}d_{L}
\end{eqnarray}
which is smaller than the box diagram contribution (35) if one remembers
(21) and (26).
Here we used the Higgs mass in (30), and $G_{F}$ is the Fermi constant and
$\alpha$ is the fine structure constant. There are diagrams other than
those in Fig.2, but their contributions are  about the same as those of
diagrams in Fig.2.

The leptonic flavor changing processes such as $K^{0}_{L} \rightarrow
e\bar{\mu}$ are also induced by the mixing of heavy fermions ,
as was noted above.
The amplitude for $K^{0}_{L} \rightarrow e\bar{\mu}$ is estimated on the
basis of diagrams analogous to those in Fig. 2, and it is given by
\begin{equation}
  \sim  (m_{W}/m_{i})^{2}(|G|/g)^{4}\alpha G_{F}
                                    \bar{s_{L}}\gamma^{\mu}d_{L}
                                    \bar{e_{L}}\gamma_{\mu}\mu_{L}
\end{equation}
This amplitude for $K^{0}_{L} \rightarrow e\bar{\mu}$  is about the same
as that for  $K^{0}-\bar{K^{0}}$ mixing in (35).

The process $b \rightarrow s\gamma$ is known to give a stringent
constraint on  vector-like schemes in general[4]. In the present
extention of the standard model, the main part of extra contributions
to $b \rightarrow s\gamma$
induced by the mixing with heavy fermions comes from the diagrams
shown in Fig.3. The amplitude in Fig.3-a is estimated at
\begin{eqnarray}
&\sim& \varepsilon e|G|^{2}(1/m_{i})\bar{s}\sigma^{\mu\nu}F_{\mu\nu}
                                    (\frac{1+\gamma_{5}}{2})b\nonumber\\
&=& \varepsilon(m_{W}^{2}/m_{i}m_{b})e|G|^{2}(m_{b}/m_{W}^{2})
                                    \bar{s}\sigma^{\mu\nu}F_{\mu\nu}
                                    (\frac{1+\gamma_{5}}{2})b\nonumber\\
&\simeq& \varepsilon (m_{W}^{2}/m_{i}m_{b})(|G|/g)^{2}eG_{F}m_{b}
                                    \bar{s}\sigma^{\mu\nu}F_{\mu\nu}
                                    (\frac{1+\gamma_{5}}{2})b
\end{eqnarray}
up to a numerical coefficient such as $1/16\pi^{2}$.
This amplitude is about the same order as the standard model prediction
\begin{equation}
\sim V_{tb}V_{ts}^\ast eG_{F}m_{b}\bar{s}\sigma^{\mu\nu}F_{\mu\nu}
                                      (\frac{1+\gamma_{5}}{2})b
\end{equation}
if one uses the upper bound $\varepsilon = 1/50$ in (26).
The contribution of Fig.3-b is obtained if one replaces $\varepsilon$
by $m_{b}/m_{i}$ in (38). We thus see that the flavor
changing radiative decay induced by the Higgs particle and heavy fermions
does not spoil the agreement of the standard model with the recent
CLEO experiment. See Ref.[4] and references therein.

{}From the above analyses, in particular , the mass correction in (33)
,which should be smaller than $\sim 1 MeV$,
and the $K^{0}-\bar{K^{0}}$ mixing in (35), which requires
$(m_{W}/m_{i})^{2}(|G|/g)^{4} \leq 10^{-8}$,
 we learn that the Higgs
coupling between a light fermion (except possibly for the top quark)
and a heavier fermion in the present model is constrained to be
\begin{eqnarray}
   (1/2)|G|v/m_{W} \leq 1/10
\end{eqnarray}
or
\begin{eqnarray}
   |G|/g \leq 1/10, \nonumber\\
   \varepsilon \leq 1/500
\end{eqnarray}
if one chooses the on-set of the  heavy fermion mass as in (25).
This (41) is the only
fine tuning in the present scheme\footnote{The (technical) problem as
to the treatment of the quadratic divergence of the Higgs mass still
remains.}.
 (The Higgs coupling among heavy fermions
can be as large as in (21)).
One can of course retain the "natural"
bound (21) if one chooses the on-set of  heavier fermion mass spectrum
at values larger than in (25) ; this choice is however less interesting
from a view point of possible physics in the near future.

The decay rate of $K_{L}^{0}\rightarrow e\bar{\mu}$ is then estimated at the
order
\begin{equation}
\Gamma(K_{L}^{0}\rightarrow e\bar{\mu})\leq 10^{-8}\times
\Gamma(K_{L}^{0}\rightarrow \mu\bar{\mu})
\end{equation}
where $\Gamma(K_{L}^{0}\rightarrow \mu\bar{\mu})$ is given by the
standard model.

It is confirmed that
$CP$ violation does not appear in the zeroth order
approximation without the Higgs coupling (see eq.(12))
and it arises solely in the Higgs sector(20);the pattern of
CP violation becomes more involved than in the standard model[1] and
CP phase is no more limited to W couplings.
(The complex mass matrix in (7), when one diagonalizes it, may
generally induce CP
phase into (20)). When one incorporates QCD,
the strong CP problem still appears in connection with
the Higgs coupling (20).

As for the neutrinos, the first three neutrinos will remain
massless if one assumes the absence of
right-handed components(i.e., if $\phi$ in (13) is a singlet)\footnote{
To be precise, the radiative correction by one-loop W-boson exchange
induces a mixing of a massless neutrino $\nu$ with a massive
neutrino $L$ of the
order of
$ \sim \alpha((|G|v)^{2}/m_{i})\bar{\nu_{L}}L_{R}$.
If one truncates the number of massive neutrinos at any finite number,
however, the neutrino $\nu_{L}$ stays massless.}.
We however expect the appearance
of vector-like heavy neutrinos above TeV region.  Cosmological
implications of those massive neutrinos may be interesting.

\section{Conclusion}

\par
Motivated by the works in Refs.[4] and [8],
 we discussed the possible existence of heavier quarks and
leptons
above 1 TeV.
Our understanding of gauge symmetry and gauge interactions is
substantial, but our understanding of matter sector is rather meager.
It may then be sensible to take $SU(2)\times U(1)$ as a guiding principle
and analyze the possible existence of heavier fermions.
By this way we can largely avoid arbitrariness in the extension of the
standard model.
A vector-like extension of the standard model examined in this note is
natural in the sense that the
validity of perturbation theory (21), combined with a fine
tuning of the Higgs coupling associated with light fermions as in (40),
 and a sensible choice of heavy
 fermion mass scale (25) lead to consistent results as a first order
approximation. A more precise analysis , which incorporates the effects
of many heavy fermions, will not alter the main features of our
semi-quantitative analysis if the spectrum of heavier fermions is
distributed rather sparsely. A moral drawn from the analysis in this
note is that flavor-changing processes ( and CP violating processes)
provide a sensitive probe of new physics beyond the standard model.

The present model as it stands is, however, a  completely
phenomenological
one:The appearance of many fermions with vector-like couplings might
be natural from some kind of composite picture of fermions, or if the
fermions are elementary their masses might  arise from a topological
origin as is suggested by (9) or from some kind of space-time
compactification. But the fundamental issue of the   breaking mechanism of
chiral and parity symmetries in (9) is not explained(see, however,Ref.[9]),
and  a picture of
unification of interactions is missing.
The breaking of asymptotic freedom of QCD by heavy fermions becomes
appreciable only at the mass scale of these heavy fermions due to
the decoupling phenomenon.

At the next generation of accelerators, we will be able to see whether
a drastic extension of the standard model such as a SUSY generalization
is realized ,or  more conventional schemes such as the one analyzed
here is realized; or we may simply find a desert as is suggested by some
unification schemes.
Clearly physics above 1 TeV
region is fascinating, and it awaits more imagination and insight.

I thank A. Yamada for critical comments.

\section*{Figure Captions}

\begin{itemize}
\item[Fig.1]
Corrections to light fermion masses. Notational conventions
follow those of eq.(27).

\item[Fig.2]
Extra diadrams for $K^{0} -  \bar{K^{0}}$
mixing. Notational conventions follow those of eq.(27).

\item[Fig.3]
Extra diagrams for flavor changing radiative decay such as
$b\rightarrow s\gamma$ decay. Notational
conventions follow those of eq.(27).

\end{itemize}

\end{document}